\documentclass[12pt,preprint]{aastex}
\usepackage{epsfig}
\begin{document}

\title{On the structure of line-driven winds near black holes}

\author{A.V. Dorodnitsyn}
\affil{Space Research Institute}
\affil{Profsoyuznaya st. 84/32, Moscow, Russia}
\email{dora@mx.iki.rssi.ru}
\and
\author{I.D. Novikov}
\affil{Niels Bohr Institute, Blegdamsvej 17, DK-2100, Copenhagen, Denmark and}
\affil{Astro Space Center of P.N.Lebedev Physical Institute  }
\affil{Profsoyuznaya st. 84/32, Moscow, Russia}
\email{novikov@nbi.dk}

\begin{abstract}
A general physical mechanism of the formation of line-driven winds at the vicinity
of strong gravitational field sources is investigated in the frame of General
Relativity. We argue that gravitational redshifting should be taken into account to
model such outflows. The generalization of the Sobolev approximation in the frame of General Relativity
is presented. We consider all processes in the metric of a nonrotating
(Schwarzschild) black hole. The radiation force that is due to absorbtion of the radiation flux in lines
is derived. It is demonstrated that if gravitational redshifting is taken into
account, the radiation force becomes a function of the local velocity gradient (as
in the standard line-driven wind theory) and the gradient of $g_{00}$.
We derive a general relativistic equation of motion describing such flow.
A solution of the equation of motion is obtained and confronted with
that obtained from the Castor, Abbott \& Klein (CAK) theory. It is shown that the
proposed mechanism could have an important contribution to the formation of
line-driven outflows from compact objects.
\end{abstract}

\section* {Introduction}
It has been demonstrated both theoretically and observationally
that accretion disks around compact objects can be powerful sources of fast
plasma outflows. Among the most important processes known to work are
magnetic and radiation driving. In fact radiation-driven winds can exist in most of systems where accretion disk
can produce enough $\rm UV$ radiation (the standard line-driven wind theory gives an
approximate value of
$L_{UV}>10^{-4} L_{edd}$). However, this conclusion should be treated with care
because of the physical conditions in a disk wind which are very different from that
of an O-type star wind. While realistic accretion disk winds are most likely driven
by the combination of the radiation and magnetic forces here we focus on the
scenario when momentum is
extracted most efficiently due to absorption of the radiation flux in lines of
abundant elements.

In the paper of Dorodnitsyn (2003) (hereafter {\bf D1}) it was proposed a mechanism
when
line-driven acceleration occurs in the vicinity of compact object so that the the gravitational redshifting can play an important role. The generalization
of these studies in the frame of General Relativity (GR) is
the problem that we address in this paper.
A mechanism that we study is quite general and can be
considered to work in any case when there is enough radiation to accelerate plasma and
radiation driving occurs in strong gravitational field.
Particularly we discuss
winds in active galactic nuclei as they manifests most important properties of
accretion disk + wind systems keeping in mind however that our treatment allows to consider
their low mass counterparts.

It is widely accepted that a supermassive black hole (BH) lies in the cores of most of active
galactic nuclei (AGN). The accretion activity around such a black hole results in a production
of a powerful continuum radiation - a defining characteristic feature of the quasar phenomenon.
The dynamical
role of this radiation is so high that it is probably responsible for the formation of fast winds
which are observed in AGN. The radiation pressure on lines plays the crucial role in acceleration
of such outflows.
The most prominent feature seen in about $10\%$ of quasar spectra
are the broad absorbtion line systems BALs - the blue-shifted UV resonance lines from highly
ionized species ($\rm N\,V,\, C\,IV, \, Si\,IV$). These come from ions of differing excitation with
bulk velocities of up to $0.2\, {\rm c}$. A successful model must also explain
a simultaneous existence of NALs - narrow absorbtion line systems ($\rm N\,V,\,C\,IV$),
seen in UV and X-rays from about a half of Seyfert galaxies and associated
with outflows of 1000 $\rm km\,s^{-1}$,
and BEL - broad emission lines present in all AGNs indicating flows as fast as
$5\cdot 10^3\,\rm km\,s^{-1}$. These well established features together with
total luminosity of up to $L\sim 10^{46}\, \rm erg \, s^{-1}$
gives us the crucial
evidence of the dynamical importance of the line-driven mechanism in AGNs.
A quasi-1D model of the quasar wind was developed in Murray et al. (1995) (however
it is not clear how justified is the assumption that equations in radial and polar
directions could be solved separately). The 2D calculations of the accretion disk
powered winds were made in
Proga, Stone \& Drew (1998) and Proga, Stone \& Kallman (2000), while the winds from
massive X-ray binaries (together with ionizing effects of the radiation from the
central source)
have been considered in Stevens \& Kallman (1990).

In the pioneering paper of Sobolev (1960), it was recognized that the problem of the radiation
transfer in lines in a continuously accelerating medium is simplified drastically in comparison
with that of a static case.
In the paper by Lucy \& Solomon (1970) it was pointed out the
importance of the line opacity for the formation of winds from hot stars.
Most of our understanding of the line-driven mechanism
is based on
the prominent paper of Castor, Abbott and Klein (1975), (hereafter CAK)
where a theory of the O-type star wind was developed.
These studies explained how a star
that radiates only a tiny fraction of its Eddington limit, can have a very strong wind. CAK was
able to demonstrate that the radiation force from en ensemble of optically thin and optically thick lines
can be parameterized in terms of the local velocity gradient. This elegant theory was further
developed
in papers of many authors.  All this work resulted in what is usually
called a "standard line-driven wind theory" (hereafter SLDW).

It is rather problematic, however, to apply directly the CAK theory to accretion disk
winds because of the
geometrical difference and because of the different properties of the spectrum emitted by the
central source of the continuum radiation.
For example, a wind in AGN is likely to be launched from
accretion disk-like structure and thus is intrinsically two dimensional with the geometry that
is close to axial symmetry. The second crucial difference is that in active galactic nuclei a
wind is exposed to a hard UV and X-ray continuum radiation that stripes electrons from abundant
elements much more effectively than the quasi-black-body radiation of the hot luminous star.
In case of AGN the radiation flux produces highly ionized species over much of the wind. The
considerable lack of the atomic data for highly ionized ionic species
and of
intrinsically 2D radiation-hydro and transfer calculations
makes a task of the realistic
modelling of AGNs winds very problematic.

In the standard line-driven wind theory a given parcel of gas sees
the matter that is upstream redshifted because of the difference in velocities
(assuming that a
wind is accelerating gradually). This helps a line to shift from the shadow produced by the
underlying matter and to expose itself to the unattenuated continuum.
It was shown in {\bf D1}, that together with Sobolev effect the gravitational redshifting
of the photon's frequency should be taken into account when calculating the radiation force.
In case of strong
gravitational field the gradient of the gravitational potential works in the same
fashion as the velocity gradient does when only Sobolev effect is taken into
account, so that the radiation force becomes
$g_l \sim(dv/dr+\frac{1}{c}d\phi/dr)$. As it was shown in {\bf D1} now the gravitational field works in exposing
the wind to unattenuated
radiation of the central source. Thus we call such a flow "Gravitationally Exposed
Flow" (GEF).

Conditions present in the inner parts of the realistic accretion disk wind
are far from being clear. However it is known that most of the radiation flux is
produced in the innermost parts of the accretion disk.
We may expect that some part of the wind which is located beyond few tens of
gravitational radii may be moving quasi-radially. To make our treatment as general
as possible, we consider spherically - symmetrical
radiationally accelerated wind. Lines and electron scattering are assumed to be the
only sources of opacity, no ionization balance is calculated.

In {\bf D1} gravitational field was considered by means of the gravitational
potential. Thus all equations were in fact derived in the flat space-time, and only the
effect of the gravitational redshifting was taken into account when calculating the
radiation pressure term. The resultant solution was then compared with CAK wind
solution. This approach is not self-consistent. In GR when calculating the radiation
force, the effect that is due to Doppler shifting should be taken into account
simultaneously with gravitational shifting (no bending of photon trajectories
is considered since the force is calculated in radial streaming limit). Obviously,
the CAK - type solution can exist only in the flat spacetime.
It is important to note that the {\it self-consistent modelling} of GEF is possible
only via general relativistic treatment.
Thus, the main
goal of this paper is to compare the general relativistic GEF solution with the SLDW
solution, obtained in the Newtonian gravity.

Here we solve GR equations of motion for radiatively accelerated wind and
calculate the radiation pressure in
the radial streaming limit in the Sobolev approximation. Making use of the Sobolev
approximation allows us not to treat the General Relativistic radiative transfer
formalism. There exist an extensive
literature, where the radiative transfer problem in GR is considered for the purpose of
hydrodynamical calculations of spherically symmetric accretion (e.g. Turolla \& Nobili 1988;
Thorne, Flammang \&  ${\dot{\rm Z}}$ytkow 1981; Nobili, Turolla, \& Zampieri 1991).
In these papers the radiative moment formalism of Thorne (1981) had been extensively used.
However, for our purposes it is not needed to use this sophisticated formalism.
In the Sobolev approximation the flow is treated in fact as non-relativistic and the
only source of the opacity is the line and electron scattering. In such an approach
it is possible to derive the radiation force without an explicit solution of the
radiative transfer equation.
Thus we use only escape probability arguments - exactly as the radiation force is
derived in the standard
line-driven wind theory (e.g. Mihalas 1978; Lamers \& Cassinelli 1999)

The plan of this paper is as follows. In Section 1 the General relativistic
equations, describing the matter that is interacting with the radiation field are
derived. Then, in Section 2 the optical depth and the radiation force in Sobolev
approximation are obtained. We obtain the equation of motion for the line-driven
wind in GR and then numerically calculate a set of its solutions in Section 3.
The results are summarized and the perspectives are discussed in the conclusions.

\section {The radiation-driven wind}
It was suggested in {\bf D1} that taking into account
the gravitational redshifting in
modelling the line-driven winds can substantially increase the efficiency of the
mechanism.

The wind is assumed to blow in the background metric described by the familiar Schwarzschild
line element
\begin{equation}\label{metrics}
ds^2=-h\,dt^2+\frac{dr^2}{\displaystyle
h}+r^2d\Omega^2\mbox{,}
\end{equation}

\noindent
where $h\equiv \left(1-\frac{r_g}{r}\right)$, $r_g=2M$, where $M$ is a black hole
mass.
Geometrized units ($c=G=1$) are used throughout this Section.
Let $\hat{{\rm O}}$ be an observer that is at rest at $r$,
that measures physical quantities like
$\hat{v}$-velocity,
$\hat{dl}= dr /\sqrt{h}$ - displacement
etc. The observer $\hat{{\rm O}}$ has the following tetrad of orthonormal basis
vectors: ${\displaystyle  \vec{e}_{\hat{t}}=h^{-1/2}\,\vec{e}_{t}}$,
${\displaystyle  \vec{e}_{\hat{r}}=h^{1/2}\,\vec{e}_{r}}$,
${\displaystyle  \vec{e}_{\hat{\theta}}=r^{-1}\,\vec{e}_{\theta}}$,
${\displaystyle  \vec{e}_{\hat{\phi}}=(r \sin\theta)^{-1}\,\vec{e}_{\phi}}$.

\noindent
The stress-energy tensor for the ideal gas reads:
\begin{equation}\label{Tij}
  T^{\alpha\beta}=(P+\rho)u^\alpha u^\beta+P
  g^{\alpha\beta}\mbox{,}
\end{equation}
where for the total mass-energy density we have

\begin{equation}\label{density}
  \rho=\rho_0(1+E_i)\mbox{,}
\end{equation}
where $\rho_0=m_b n_b$ is the barionic rest-mass density, $n_b$ is the baryon
number density, and $E_i$ is
the internal energy of the ideal gas per unit mass.

The continuity equation reads:

\begin{equation}\label{contin}
(n_b u^\alpha)_{;\alpha}=0\mbox{.}
\end{equation}
The process of the interaction of the matter with the radiation is
described by the four-force density which is given by
\begin{equation}\label{rforce}
  G^\alpha=\int_0^\infty \, d\nu \oint d\Omega
  \left( \chi_\nu I_\nu-\eta_\nu \right) n^\alpha \mbox{,}
\end{equation}
where $\chi_\nu$ is the opacity and $\eta_\nu$ is the emissivity.
$I_\nu$ is the specific intensity of photons of the frequency $\nu$,
propagating in the direction $n^\alpha$, the space component $G^i$ gives
the net rate of the radiative momentum input, while $c\,G^0$ equals the
rate of the radiative energy input.

The equations of hydrodynamics for the matter interacting with the radiation filed
read (see, e.g., Mihalas \& Mihalas, 1984):

\begin{equation}\label{TGij}
  T^{\alpha\beta}_{;\beta}-G^\alpha=0\mbox{.}
\end{equation}
Applying the projection tensor

\begin{equation}\label{project}
  P_{\alpha\mu}\equiv g_{\alpha\mu}+u_\alpha u_\mu
\end{equation}
to the equation (\ref{TGij}) will result in the Euler equation of
motion:

\begin{equation}\label{Euler1}
P_{\alpha\mu}\left\{T^{\alpha\beta}_{;\beta}-G^\alpha\right\}=0\mbox{.}
\end{equation}
Calculating equation ~(\ref{Euler1}) in the orthonormal frame of ${\hat O}$ we
obtain:

\begin{equation}\label{Euler2}
 \frac{P+\rho}{h}\left( h\gamma^2 v \frac{dv}{dr}+\frac{M}{r^2}\right)+\frac{dP}{dr}-
 \frac{1}{\sqrt{h}}(G^{\hat 1}-v G^{\hat 0})=0\mbox{,}
\end{equation}
where $\gamma=(1-v^2)^{-1/2}$ and it was taken into account that the radiation
force measured by ${\hat O}$
is given by:
$G^{\hat 1}=G^1/\sqrt{h}$, $G^{\hat 0}=\sqrt{h}\,G^0 $.
These physical components of
the radiation four-force density can be further transformed
to the frame that instantly coincides with the moving gas by making use of the
corresponding Lorentz
transformation: $\gamma(G^{\hat 1}-v\,G^{\hat 0})=G^{\hat 1}_0$.
The relative importance of the frame-dependent term  $G^0$ in ~(\ref{Euler2})
is $O(v/c)$ relative to $G^1$ (Mihalas \& Mihalas 1984).
The radiation force $G^{\hat 1}_0=G^{\hat 1}_{0,\,c}+G^{\hat 1}_{0,\,L}$ is the
sum of a radiation force due to absorbtion in
continuum and a force which results from the line
transition, as measured by the physical co-moving observer.
The latter should be calculated taking into
account both the shifting of frequency due to Doppler effect and
the gravitational redshifting. Note that equation (\ref{Euler2}) corresponds
to equation [11] of Nobili, Turolla, \& Zampieri (1991), if to imply that $G^{\hat 1}$
consists only of the part of the force that is due to continuum radiation flux.
From equation (\ref{contin}) we obtain continuity equation in the form:

\begin{equation}\label{contin2}
  \rho_0 v r^2 \gamma \sqrt{h}=\frac{\dot M}{4\pi}\mbox{,}
\end{equation}
where ($\gamma\rho_0$) - is the barionic rest-mass density as measured in the laboratory
frame.
\newline The key ingredient of the CAK theory is that the optical depth which is due to the
line absorbtion can be expressed as a function of the velocity
gradient in the flow. The same is true in the case when a photon, emitted somewhere deeply in the
potential well, will become resonant with the line absorption both due to velocity
difference and due to GR effect of redshifting.

\section{Optical depth and radiation force in Sobolev approximation}
A photon emitted at a
given radius will suffer a continuous both
gravitational and Doppler redshifting and may become resonant with a line transition at
some point
downstream. Thus a ray of a frequency $\nu_d$, emitted by the matter that for simplicity is assumed
to be at rest at radius $r_d$, at a given point $r$ has a frequency $\nu_{lab}$, as
measured by the observer ${\hat O}$ and that is obtained from relation
(see, e.g. Landau \& Lifshitz 1960):

\begin{equation}\label{nuinv}
\nu_{d}\sqrt{h_d}=\nu_{lab}\sqrt{h(r)}=\nu^{\infty}\mbox{,}
\end{equation}
where $\nu^\infty$ is the frequency of the ray at infinity.
We restrict ourself to the {\it radially streaming photons}
only and assume that they are emitted from a point source. In
such a case the Sobolev optical depth may be calculated without
solving the radiation transfer equation. Optical depth between $r_d$ and  a given
point $r$
can be written  (Novikov \& Thorne 1973):

\begin{equation}\label{tau}
  \tau_l=\int_{r_{d}}^r \chi_{l,\, lab} \,\hat{dl}=
  \int_{r_{d}}^r \chi_{l,\, lab} \,\frac{dr}{(1-\frac{2M}{r})^{1/2}}\mbox{,}
\end{equation}
where $\hat dl$ is the proper-length element, and $\chi_{l,\, lab} \,(\rm cm^{-1})$ is the absorbtion
coefficient as measured by ${\hat O}$. However, it is more appropriate to measure
absorbtion (as well as the emission) in the local frame co-moving with the fluid.
In such a case, opacity is transformed according to the reation:
$\chi_{l,\, lab}=\chi_{l,com}\,\tilde \nu/\nu_{lab}$, where
in the co-moving frame the frequency of the ray is
${\tilde\nu}=\gamma\nu_{lab}(1-\beta)$, $\beta=v/c$, and $\chi_{l,com}$ is the absorbtion
coefficient measured by the co-moving observer. In the co-moving frame, the line-center
opacity is determined by the following relation:

\begin{equation}\label{kappa}
  \chi^0_l=\frac{\pi e^2}{m c} g f
  \frac{N_L/g_L-N_U/g_U}{\Delta\nu_D}\mbox{,}
\end{equation}
where $\Delta\nu_D=\nu_0 v_{th}/c$ is the Doppler width, and $\nu_0$ is the
line frequency, $f$ is the oscillator strength of the transition, $g$ is the
statistical weight of the state,
$N_U$, $N_L$ and $g_U$, $g_L$ are respective populatios and
statistical weights of the corresponding levels of the line transition.

Then, one have to compare ${\tilde \nu}$ with the frequency of the
line $\nu_0$, to see whether it is in the range of the line profile
$\varphi(\tilde\nu-\nu_0)$. That is in fact a standard procedure
that is followed in order to calculate the Sobolev
optical depth, but with an additional step - the gravitational
redshifting.
Thus, both Doppler and gravitational redshifting of the
photons frequency are taken into account.

\noindent
Introducing a new frequency variable:

\begin{equation}\label{xdef}
y \equiv \tilde\nu-\nu_0=\gamma (1-\beta)\frac{\nu^\infty}{\sqrt{h}}-\nu_0\mbox{,}
\end{equation}
We change the integration variable in ~(\ref{tau}) from $r$ to $y$:

\begin{equation}\label{tau1}
  \tau_l=\int_{r_d}^r \frac{\tilde\nu}{\nu_{lab}}\varphi(\tilde\nu-\nu_0)\chi^0_l \Delta\nu_D
  \,\frac{dr}{\sqrt{h}}=\int_{y(r_d)}^{y(r)} \frac{\tilde\nu}{\nu_{lab}}
  \frac{\varphi(y) \chi^0_{l}\Delta\nu_D }
  {\nu^\infty\left(\displaystyle \frac{d\eta}{dr}-
  {\rm w}\frac{\eta}{\sqrt{h}}\right)}
  \,dy \mbox{,}
\end{equation}
where ${\displaystyle \eta\equiv\gamma(1-\beta)}$. The relation
~(\ref{xdef}) was used in order to calculate $dx/dr$. The last
term in the denominator of ~(\ref{tau1}) is due to the "gradient
of the gravitational field": $\displaystyle {\rm
w}=\frac{d}{dr}\sqrt{h}$. Note that $\displaystyle {\rm w} \cdot
c^2\equiv\frac{GM}{r^2 \sqrt{h}}$ - equals the acceleration of the
free-falling particle that was initially at rest in the
Schwarzschild metric. In case of $w=0$ we obtain the result of
Hutsemekers \& Surdej (1990): $\displaystyle \tau_l=\frac{\chi^0_l
v_{th} (1-\beta)}{\gamma dv/dr}$. Assuming that the line profile
is a $\delta$- function, or, equivalently, that the region of
interaction is infinitely narrow, we find the optical depth in
Sobolev approximation: ${\displaystyle
  \tau^\star_l=\frac{\chi^0_l\Delta\nu_D(1-\beta)}{\displaystyle \nu_0\left|
  \sqrt{h}\gamma\frac{d\beta}{dr}+{\rm w}\gamma^{-1}\right|}
  }$,
where it was taken into account that
$\eta'=-\gamma^3(1-\beta)\beta'$. In our treatment we retain only
terms $O(v/c)$ (in the equation of motion) and thus resultant
Sobolev optical depth can be written in the form:

\begin{equation}\label{tau2}
\tau_l =\frac{\kappa^0_l \rho_0 v_{th}}{\displaystyle \left| \sqrt{h}\frac{dv}{dr}+
  c{\rm w}\right|}\mbox{,}
\end{equation}
where $\kappa^0_l$ ($\rm cm^2\cdot g^{-1}$) is the mass absorbtion
coefficient: $\kappa^0_{l}=\chi^0_l/\rho_0$, measured in the rest-frame of the fluid,
and $\gamma=1$.

In the weak field limit the optical depth ~(\ref{tau2}) will transform to equation (11) of {\bf D1}. If there
is no gravitational redshifting taken into account then the Sobolev optical depth is obtained:
${\displaystyle \tau_{sob}=\frac{\kappa^0_l\rho_0 v_{th}}{\displaystyle \left| dv/dr \right|}
  =\chi^0_l\,l_{sob}}$,
and the Sobolev length scale $l_{sob}=v_{th}/\left(dv/dr\right)$
determines a typical length on which a line is shifted on about its thermal width.
In general case from ~(\ref{tau2}) we conclude that

\begin{equation}\label{tauSob}
  \tau_l=\chi^0_l\,l_{GEF}\mbox{,}
\end{equation}
where ${\displaystyle{l_{GEF}}=\frac{v_{th}}{\displaystyle \sqrt{h}\frac{dv}{dr}+
  c{\rm w}}}\mbox{.}$

The radiation force form a single line exerted by the material as measured
in its rest frame reads:

\begin{equation}\label{rforce1}
  g_i  \simeq \frac{F^0_\nu (\nu_l)\chi^0_l/\rho_0\Delta\nu_D}{c} \frac{1-e^{-\tau_l}}{\tau_l}
  \mbox{,}
\end{equation}
where $F^0_\nu (\nu_l)(\rm erg \,\cdot cm^{-2} \,\cdot Hz^{-1} \,\cdot s^{-1})$
is the radiation flux at the line frequency in the rest-frame of the fluid.
Note that $\tau_l$ in ~(\ref{rforce1}) should be calculated with taking into account
the redshifting and is given by ~(\ref{tau2}).
 The $e^{-\tau_l}$ term reflects the fact that the incident
flux at $r$ is reduced in comparison with the initial flux $F_\nu$. $1-e^{-\tau_l}$
gives the "penetration probability" for a ray to reach a given point.

In our treatment we neglect special relativistic terms all equations.
We may expect that final results will be at least qualitatively correct
for the flows as fast as $\sim 0.2 \div 0.3c$.
From ~(\ref{rforce1}), we conclude that if a line is optically thin, the radiation
force does not depend upon the redshifting law ~(\ref{nuinv}):
$\displaystyle g_{\rm thin}=\frac{\chi_l\Delta\nu_D}{c\rho_0} F_\nu $. On the other hand, when
$\tau_l>>1$
the optically thick line produces force that can be roughly described by the following:
$\displaystyle g_{\rm thick}=\frac{\chi_l\Delta\nu_D}{c\rho_0} \frac{F_\nu}{\tau}=
\tau^{-1} g_{\rm thin}$, and thus it is independent of the line strength and $g_l\sim dv/dr$.

The difference between $\rho_{0,lab}$ and $\rho_0$ is $O(v^2/c^2)$.
Note, that in our case $g^0_{\rm thin}=\chi^0_l F^0_\nu/c=g^{lab}_{\rm thin}+O(v^2/c^2)$,
$g^0_{\rm thick}=F^0_\nu/c=g^{lab}_{\rm thick}+O(v^2/c^2)$, and
a factor ($1-\beta$) was omitted when calculating ~(\ref{tau2}).

According to the well accepted notation let introduce the optical depth parameter:

\begin{equation}\label{t1}
t=\frac{\sigma_e \, \rho_0 v_{th}}{\displaystyle \left| \sqrt{h}\frac{dv}{dr}+{\rm c w}\right|}\mbox{,}
\end{equation}
where $\sigma_e$ is the electron scattering opacity per unit mass, and
$t$ is connected to $\tau_l$ via relation: $\tau_l=\xi t$, where
$\xi=\kappa^0_l/\sigma_e$.

The role of the parameter $\xi$ is very important because it allows to separate
the line optical depth into two parts: the first ($\xi$) that depends on statistical
equilibrium, and the second ($t$) which depends only on the redshifting law
~(\ref{xdef}). It will alow us to use the standard parameterization law for the force
multiplier when calculating the radiation force.
Summing ~(\ref{rforce1}) over the ensemble of of optically thin and optically thick lines
we obtain total radiation acceleration:

\begin{equation}\label{g1}
  g_l=\sum_l g_i=\frac{F\sigma_e}{c}\, M(t)\mbox{,}
\end{equation}
where $F$ is the total flux and the "force multiplier" M(t) equals:

\begin{equation}
  M(t)=\sum_{\tau_l<1}\frac{F_\nu}{F}\, \xi \, \Delta\nu_D+
  \sum_{\tau_l>1}\frac{F_\nu}{F} \frac{\Delta\nu_D}{t}\mbox{.}
\end{equation}
CAK found that M(t) can be fitted by the power law:

\begin{equation}\label{g3}
M(t)=kt^{-\alpha}\mbox{.}
\end{equation}

\subsection* {The equation of motion}
Combining equations (\ref{t1}), (\ref{g1}), (\ref{g3}) and the equation of continuity
(\ref{contin2}), the equation of motion describing a stationary,
spherically-symmetric, wind can be cast in the form:

\begin{eqnarray}\label{Euler3}
&& \frac{b}{h}\left(v h
\frac{dv}{dr}+\frac{GM}{r^2}\right)+\frac{1}{\rho_0}\frac{dP}{dr}-
  \frac{\sigma_e}{\sqrt{h}}\frac{L}{4\pi r^2 c}\nonumber\\
&& -\frac{\sigma_e}{\sqrt{h}}\frac{L}{4\pi r^2 c} k
  \left(\frac{4\pi}{\sigma v_{th} {\dot M}} \right)^\alpha
  \left\{\sqrt{h}\,vr^2\left[\sqrt{h}\frac{dv}{dr}
  +c {\rm w}\right]\right\}^\alpha=0\mbox{,}
\end{eqnarray}
We adopt the equation of state for the ideal gas: $P=\rho_0{\cal
R}T$, $E_i=3/2\,{\cal R}T$, where ${\cal R}=k/m_p$ is the gas
constant, $L$ is the luminosity, ${\displaystyle
b(T)=\frac{\rho+P/c^2}{\rho_0}}\simeq1$ since for the conditions typical
for any line-driven wind, quantities $P/c^2$, $E_i \rho_0/c^2$ are
vanishingly small and $\rho=\rho_0+O(v^2/c^2)$. Furthermore we
assume that the gas is isothermal and hence:

\begin{equation}\label{eq_state}
P=a^2\rho_0\mbox{,}
\end{equation}
where
$\displaystyle a^2=\left(\frac{\partial P}{\partial
\rho_0}\right)_T={\cal R}T$.
Using the continuity equation (\ref{contin2}) and the equation of state (\ref{eq_state}) to transform the
$dP/dr$ term,
after some manipulation, from (\ref{Euler3}) we obtain:
\begin{eqnarray}\label{Euler31}
&&\left(1-\frac{v^2_s}{v^2}\right)v\frac{dv}{dr}+\frac{GM}{r^2\sqrt{h}}
\left(\frac{1}{\sqrt{h}}-\Gamma\right)
-a^2\left(\frac{2}{r}+\frac{\rm w}{\sqrt{h}}\right)\nonumber\\
&&-\frac{\sigma_e}{\sqrt{h}}\frac{L}{4\pi r^2} k
  \left(\frac{4\pi}{\sigma_e v_{th} {\dot M}} \right)^\alpha
  \left\{vr^2\sqrt{h}\left[\sqrt{h}\frac{dv}{dr}
  +c {\rm w}\right]\right\}^\alpha=0\mbox{,}
\end{eqnarray}
where
$\displaystyle v^2_s=\left(\frac{\partial P}{\partial
\rho}\right)_T=\frac{{\cal R} T}{b}\simeq a^2$, and $\Gamma=L\sigma_e/4\pi c G M$.

Analogously to CAK theory, equation (\ref{Euler31}) has a critical point that is not
a sonic point, which is evident from the fact that, even if $v=v_s$
the last term containing $dv/dr$ does not
vanish. Thus we are looking for a solution that starts subsonically from $r_i$,
goes smoothly through the critical point and then reaches terminal velocity $v^{\infty}$ at infinity.
Converting equation
(\ref{Euler31}) to nondimmensional units according to the
following formulas:

\begin{equation}
x=\frac{r}{r_g}\mbox{,}\qquad {\tilde v}=\frac{v}{v_c}\mbox{,}
\end{equation}
and for simplicity omitting tilde, equation (\ref{Euler31}) reads:

\begin{eqnarray}\label{Euler4}
&&F(p,v,x)\equiv\left(1-\frac{a^2_1}{v^2}\right)v
p+\frac{\zeta^2}{2  x^2\sqrt{h}}
\left(\frac{1}{\sqrt{h}}-\Gamma\right)
-a^2_1\left(\frac{2}{x}+\frac{1}{2 h x^2}\right)\nonumber\\
&&-\frac{\zeta^2}{x^2
\sqrt{h}}\mu\left\{\frac{vx^2\sqrt{h}}{\zeta^2}
\left[p\sqrt{h}+\frac{\zeta}{2\sqrt{h}x^2}\right]\right\}^\alpha=0\mbox{.}
\end{eqnarray}
where $p=dv/dx$, $\zeta=c/v_c$, $a_1=a/v_c$, and
${\displaystyle \mu=\frac12\,\Gamma k\left( \frac{8\pi GM}{{\dot M}\sigma
v_{th}}\right)^\alpha}$, which determines the rate of an outflow.
Equation (~\ref{Euler4}) is nonlinear with respect to $p$, so that
a special technique must be used. This treatment is well known and
had been used in CAK theory. Equation ~(\ref{Euler4}) may have
zero, one or two roots depending on various arguments that are
input into it. The position of the critical point is determined by
the "singularity condition":

\begin{equation}\label{sing_cond}
\left(\frac{\partial F}{\partial p}\right)_c=0\mbox{.}
\end{equation}
The velocity gradient must be continuous in the whole domain of interest
thus requiring,
that the second derivative is defined in the critical point.
The regularity condition reads:

\begin{equation}\label{reg_cond}
  \left(\frac{\partial F}{\partial x}\right)_c+
  p_c \left(\frac{\partial F}{\partial v}\right)_c=0\mbox{.}
\end{equation}
For a given position of the critical point $x_c=r_c/r_g$ there are three
parameters which should be determined: $\mu$, $p_c$, $v_c$.
The equation for the third parameter is obtained from ~(\ref{Euler4})
(calculated at the critical point $x_c$):

\begin{equation}\label{Fcr_cond}
F_c=0\mbox{.}
\end{equation}
Obtaining of these equations is straightforward but tedious and we derive them in
Appendix.

\section{Wind structure}
The procedure of solving of equations ~(\ref{Euler4})-~(\ref{Fcr_cond})
is straightforward.
For a given position of the critical point $r_c$ we can calculate the value of the
velocity and velocity gradient in the critical point.
Equations ~(\ref{sing_cond})-~(\ref{Fcr_cond}) are used to calculate $p_c$ and
$v_c=v_s/a_1$. Since $\zeta=(c/v_s)a_1$, there is only one independent parameter
$a_1$. Then adjusting the position of the critical point we integrate the equation
~(\ref{Euler4}) inward, looking for the solution that satisfies the inner
boundary condition.

To compare self-consistently SLDW and GEF solutions, they should both be matched
to a flow at rest at some given radius $r_{in}$.
When a solution for a stellar wind is calculated, a photospheric boundary
condition is usually adopted. In such a case the position of the critical point is
adjusted in order to obtain a solution that gives the position of the photosphere
$r_{ph}=r(\tau\simeq 2/3)$ at a given radius $R$, which is identified with the
radius of a star (see, e.g. Bisnovatyi-Kogan 2001). However,
in case of disk-powered winds this procedure is clearly non-physical and we should
approach a different strategy. It is not possible to fit
self-consistently a solution for the spherically-symmetric wind with that of the
accretion disk. Generally, it would require a 2D modelling which is beyond of the
current studies.

To be able to compare self-consistently the GEF solution with SLDW solution we
should start integration in deeply subsonic ($v\ll v_s$) region,
from some initial density
$\rho_{in}$. This is equivalent to the problem of the continuous fitting of the
wind solution with that of a static core when calculating a structure of a star
when mass loss is taken into account.
When a solution for a {\it stationary} outflowing stellar envelope is fitted to
that of a static core only $T$ and $\rho$ should
be matched at a fitting point $r_{in}$: $T^{env}(r_{in})=T^{core}(r_{in})$,
$\rho^{env}(r_{in})=\rho^{core}(r_{in})$ (see, e.g. Bisnovatyi-Kogan \& Dorodnitsyn 1999),
where the the fitting point is located in a deep subsonic region $v(r_{in})\ll v_s$.

We calculate the wind solution in two stages. First we find the subcritical part of
the solution. On that stage the two-boundary value problem must be solved in order
to give
the position of the critical point, that is adjusted in such
way that the solution satisfies the inner boundary condition.
For the inner boundary condition we prescribe
$\rho_{in}$ at $r_{in}$, provided, that $v(r_{in})\ll v_s$.
After some experimentation we found that the relative error is indeed becoming
vanishingly small as the velocity reduces to $v_{in}\leq 0.01 \, v_s$.
For the subcritical part
of the integration domain we have: $[\,x_{in}, x=x_c-\delta x\,]$,
$[\,v_(r_{in}), v=1-|p_c|\delta x\,]$, and $x_{in}\equiv r_{in}/r_g$, provided that
$\delta x$ is small enough. For a given position of the critical point, $p_c$ is
For a given position of the critical point $r_c$, $p_c$ and $v_c$ are
calculated from ~({\ref{App_p1}) and ~({\ref{App_poly}) of Appendix.

The critical tool of our calculations is the relaxation method with automatic
allocation of mesh points. We found that a standard approach based
on Runge-Kutta solvers is not appropriate in our case because it is very difficult to
obtain the desired accuracy when fitting to the inner boundary condition using the
shooting strategy.
There exist a wide variety of relaxation methods for the solution of BVP, for
example the method of Heney is widely used in stellar evolution calculation.
Our original code is based on the prescriptions of Press et al. (1992) and
allows to obtain the solution that satisfies the boundary condition
simultaneously adjusting the position
of the critical point. The position of the critical point $x_c$ is treated as an
additional variable, as described in Press et al. (1992),
$300$ grid points are used in the subcritical domain.

After the subcritical part of the solution is fixed we step
from the critical point outward ($v=1+|p_c|\,\delta x, x=x_c+\delta x$), and integrate
equation ~(\ref{Euler4}) to large radii obtaining $v^\infty$.

The results of the numerical integrations are present
on Fig.1 and in the Table 1, where the following notations are adopted:
$\Delta_c^{\rm CAK}=x_c^{\rm CAK}-x_{c,\rm GEF}$,
$\Delta_c^{\rm MLDW}=x_c^{\rm MLDW}-x_{\rm c,GEF}$;
$\Delta_{\rm MLDW}^\infty=(v^{\infty}_{\rm GEF}-
v^\infty_{\rm MLDW})/v^\infty_{\rm MLDW}$ and
$\Delta_{\rm CAK}^\infty=(v^{\infty}_{\rm GEF}-v^\infty_{\rm CAK})/v^\infty_{\rm CAK}$.
Each set of solutions ($s_i$) is characterized by the position
of the GEF critical point. Results on Fig.1 were obtained for a case of
$\alpha=1/2$. This case is especially suitable for integration because in that case
the equation of motion can be transformed to quadratic equation with respect to
$v'$. Curves are calculated for the following set of parameters:
$\Gamma=0.5$, $k=0.03$, $M_{BH}=10^8 M_{\odot}$,
$T=4\cdot 10^4K$.

\subsection{Standard Line-Driven Wind solution}
The properties of the CAK solution is studied in great detail and it is illustrative
to present some of these results in the notations adopted in this paper.
Thus, the equation of motion
reads:
\begin{eqnarray}
&&F(p,v,x)=\left(1-\frac{a^2_1}{v^2}\right)v p+\frac{\zeta^2}{2x^2}
\left(1-\Gamma\right)
-a^2_1 \frac{2}{x} \nonumber\\
&&-\frac{\zeta^2}{x^2}\mu\left\{\frac{vx^2}{\zeta^2}
p\right\}^\alpha=0\mbox{.}
\end{eqnarray}

In the SLDW case the velocity gradient in the critical point can be expressed
explicitly ${\displaystyle \left( \frac{dv}{dx} \right)_c=1/x_c }$. The value of
$a_1=v_s/v_c$ is found from the following relation:

\begin{equation}
\frac{1}{a_1^2}=1-\frac{\alpha}{1-\alpha}
\left\{\frac{1-\Gamma}{2x_c\epsilon^2}-2\right\}\mbox{,}
\end{equation}
where $\epsilon=v_s/c$, and all other quantaties have the same meaning as
in ~(\ref{Euler4}). After $p_c$, and $a_1$ have been found the value of $\mu$
is calculated from the following:
\begin{equation}\label{mu_cak}
\mu^{\rm CAK}=\frac{\left(1-a_1^2\right) (p_c\,x^2)^{1-\alpha}}{\alpha\,
\zeta ^{2(1+\alpha)}}
\mbox{.}
\end{equation}
As it clear from the Fig.1, the SLDW critical point is situated considerably
farther downstream then the GEF critical point, moreover in that part of the domain,
where GEF is important, $v^{\rm GEF}_c<<v^{\rm CAK}_c$. For example, for
$x^{\rm GEF}_c=50$, $x^{\rm CAK}_c=75$ and $v^{\rm CAK}_c/v^{\rm GEF}_c\simeq 37$
(c.f., {\bf D1}). The velocity in the critical point spans the following range of values:
$v_c/v_{th}=1541$ for $x_c=22.5$ drops to $v_c/v_{th}=674$ for $x_c\simeq 150$.

\subsection{Modified Line-Driven Wind solution}
It is important to understand what is the relative impact to the total dynamics,
of the acceleration that is due Sobolev effect ( when only Doppler effect is
taken into account) alone in comparison
with GEF case. In other words we would like to see what happens if we take into account only
Doppler effect, neglecting gravitational redshift, but treating the rest
of the problem in General Relativity.
As it was emphasized in the Introduction this approach is not self-consistent,
nevertheless considering such a solution can give us an
important insight to the relative importance of the effect. In such a case the
equation of motion will be identical to equation ~(\ref{Euler4}) apart of the last
term, which in such a case reads:
\begin{equation}\label{lat_term}
g^{\rm MLDW}_l\sim -\frac{\zeta^2}{bx^2 \sqrt{h}}\mu\left\{\frac{vx^2}{\zeta^2}
h p\right\}^\alpha=0\mbox{.}
\end{equation}
In {\bf D1} it was found that using the Paczynski-Wiita (PW) modified potential
results in an increase of the terminal wind velocity. It was considered to be natural,
because a wind needs to have steeper gradient in order to move out of the sharper
potential well. As it is clear from Fig. 1, making use of general relativity
increases this effect (c.f. {\bf D1}, Fig. 4). However, for a given radius, the
position of Modified Line-Driven Wind (hereafter MLDW) critical point differs only slightly from that of GEF:
$\Delta^{\rm MLDW}_c=x^{\rm MLDW}_c-x^{GEF}_c \sim 0.02 \div 0.7$ for a considered range
of $x^{GEF}_c$. Note, that for $x_c=15\div 100$, critical velocity of MLDW is almost
four times greater then in the GEF case. For example, for $x_c\simeq 15$,
$v_c/v_{th}\simeq 99$, and  $x_c\simeq 100$ $v_c/v_{th}\simeq 98$.
Thus our recent studies confirms the qualitative picture which was
obtained in {\bf D1}.

\subsection{Gravitationally Exposed Flow}
In the general relativistic calculation presented here we find
a considerable gain in the wind terminal velocity in comparison with both the CAK
solution and the semi-classical solution of {\bf D1}.

\begin{tabular}{|l||l|l|l|l|l|l|l|l}
\multicolumn{4}{c}{}\\
\hline
${\rm Model}$ & $x^{GEF}_c$ & $\Delta_c^{\rm CAK}$ & $\Delta_c^{\rm MLDW}$&
$\Delta^\infty_{\rm MLDW}$&
$\Delta^\infty_{\rm CAK}$ \\
\hline

${\rm s_1}$ &$ 15$ &$7.47$& $0.016$ &
$0.23$ & $0.57$ \\

$s_2$ & $20 $ & 9.95 & 0.03&  $0.232$  & $0.51$\\

$s_3$ & $50 $ & 25 & 0.16 & $0.24$ & $0.37$\\

$s_4$ & $100 $ & 50 & 0.66 & $0.29$ & $0.2$ \\

\hline
\end{tabular}

{Table 1: Comparison of GEF solution and SLDW solution. See text for details.}

The position of the GEF critical point is found to be closer to BH than the SLDW
critical point.
From the Table 1. we see that even when the solution originates (in fact it is
determined by the position of the critical point) sufficiently far from BH there
exist a valuable gain in $v^{\rm GEF}$ in comparison with CAK case.

\section{Discussion and conclusions}
Line-driven winds represent the most plausible explanations of fast
outflows from various astrophysical
sources.

The most successful theory presented so far has been that
of Castor, Abbott and Klein (1975), (CAK), where a theory of an O-type star wind was
developed. A
modification of this approach was used to explain outflows in active galactic nuclei.
It is now widely accepted that most of AGN manifests itself by
fast uncollimated outflows. Such fast (up to $0.1c$) winds are believed to be driven
by the radiation pressure on spectral lines. An accretion disk around supermassive
black hole (BH) is believed to be a
source of such a powerful continuum radiation.
A lot of work have been done to put together this jig-saw
puzzle (see,
e.g.
Arav \& Li  (1994);
Arav, Li \& Begelman, (1994);
Murray et al. (1995);
Proga, Stone \& Drew (1998); Proga, Stone \& Kallman (2000) and
Stevens \& Kallman (1990), note that this problem is
closely connected with that of modelling of winds form massive X-ray binaries)

It is in the paper of {\bf D1}) where the mechanism
of plasma acceleration due to absorbtion of the radiation flux in lines in a strong gravitaional
field was first investigated. A photon emitted deeply in the potential well will suffer a
continuous redshifting of frequency, that should be additionally taken into account if
a wind is accelerated near compact object.
A parcel of gas sees matter that is upstream as being
redshifted both due to difference in velocities (as in classical Sobolev
approximation) and due to gravitational redshifting which exist regardless on
whether gas is moving or not. As it had been demonstrated in {\bf D1}, the radiation pressure
on spectral lines becomes a function of the local velocity gradient and the gradient of the
gravitational potential. Thus it has been concluded that the greater the
gravitational redshifting, the more effectively a line is shifted to the extent
where the radiation flux is un-attenuated by the line opacity. Since in such a case the gravitational
field 'exposes' a wind to the un-attenuated continuum, we call this kind of flow 'gravitationally exposed flow'
(GEF).

The generalization of these studies in the frame of General Relativity (GR) is
the problem that has been addressed in this paper. Only this approach allows to take
self-consistently into account both Doppler and gravitational redshifting.
The main goal of these studies was to confront GEF regime with CAK wind.

\noindent
Using the Sobolev approximation,
generalized within the framework of GR, the acceleration
that is due to
absorbtion in a single line was found to be
\newline
${\displaystyle g^{GEF}_l \sim
(\sqrt{g_{00}}\frac{dv}{dr}+c\frac{d\sqrt{g_{00}}}{dr})}$ that should be
compared with the CAK case: ${\displaystyle g^{CAK}_l\sim \frac{dv}{dr}}$.
The acceleration due to gravitational redshifting is most important at the bottom of the
wind where the velocity is small. In our treatment terms of the order
$O(v^2/c^2)$ had been neglected, thus it is roughly accurate for velocities as fast as $\sim 0.2c$.
However the relative lost of accuracy at mildly relativistic velocities is not
very important, because our goal was to calculate the relative importance of the
effect (which is important at low $v$).
The derived general relativistic equation of motion has a critical point that is
different from that of CAK (Note that the CAK point is not a sonic point). Since
this equation is nonlinear with respect to velocity gradient the CAK approach to
such an equation was adopted. In order to compare GEF solution with the SLDW we
numerically solve the two boundary value problem. A relaxation method is found to be
a very important tool to find such a solution. An integral
impact of the gravitational redshifting to the radiation acceleration can give
a considerable gain in terminal velocity ($\Delta v^\infty/v^\infty>0.4$).
We defer a detailed study of the mathematical properties (including the analysis
of stability) of the GEF equations to
a separate paper.

Finally we summarize most important results which have been obtained in the current
studies:
\begin{enumerate}
\item
A wind driven by the radiation pressure on spectral lines was considered in the
frame of General Relativity. Following Dorodnitsyn (2003), we argue that it is
important to take into account the gravitational redshifting of the photon's
frequency, when calculating the radiation force.
\item
A generalization of the Sobolev
approximation in GR was developed and the general relativistic equation of motion
with the radiation pressure force on spectral lines
was derived.
\item
The results of the numerical integration of the equation of motion
demonstrate that taking into account gravitational redshifting can result in a wind
that is considerably more fast than  previously assumed on the ground of the CAK
theory.
\end{enumerate}

\subsection*{Acknowledgments:} This work was supported in part by RFBR grant
N 020216900, and INTAS grant N 00491, and the Russian Federation President grant
MK-1817.2003.02,
and Danmarks Grundforskningsfond through its support for establishing of the
Theoretical Astrophysics Center, and by the Danish SNF Grant 21-03-0336.
Authors are grateful to Bosnovatyi-Kogan G.S., for stimulating and helpful discussions.
A.D. thanks M. Lyutikov for useful discussion, A.D. also thanks the Theoretical Astrophysics Center
for hospitality during his visit. We also would like to thank the anonymous referee for wise comments.

\subsection*{Appendix}
From the 'singularity' condition ~({\ref{sing_cond}) we obtain the value of $\mu$:

\begin{equation}\label{App_mu1}
\mu =\frac{(1 - a^2_1) (2 h_c p_c x_c^2 + \zeta)}{2\alpha\zeta^2
\, \sqrt{h_c}}
  \left(\frac{x^2_c \sqrt{h_c} }{\zeta^2}(p_c\sqrt{h_c}+
\frac{\zeta}{2x_c^2\sqrt{h_c}} ) \right)^{-\alpha}\mbox{,}
\end{equation}
where $h_c\equiv h(x_c)$
Substituting ~({\ref{App_mu1}) to ~({\ref{Fcr_cond}) and
solving the resultant equation with respect to $p_c$ will result:

\begin{equation}\label{App_p1}
p_c=\frac{-{{a_{1}}}^2\,\left( \alpha  + 4\,{h_c}\,{x_c}\,\alpha -
\,\zeta  \right)  +
     \zeta \,\left( -1  + \alpha \,\zeta  - {\sqrt{{h_c}}}\,\alpha \,\zeta \,\Gamma  \right) }{2\,
     \left( {{a_{1}}}^2 -1\right) \,{h_c}\,
     {{x_c}}^2\,\left(\alpha -1 \right) }\mbox{.}
\end{equation}
After a lengthy calculus of the 'regularity condition' ~({\ref{reg_cond}), taking into
account ~({\ref{App_mu1}) the quadratic equation can be obtained:

\begin{equation}\label{App_poly}
c_2 p_c^2+ c_1 p_c+c_0=0\mbox{,}
\end{equation}
where $c_i$ should be calculated from the following:

\begin{eqnarray*}
c_0&=&-\left( {{a_{1}}}^2\,\left( 8\,{{h_c}}^2\,\alpha  -
2\,{h_{1c}}\,\alpha  + {h_{1c}}\,\,\zeta  \right)
    \right)\nonumber\\
    &&+\,{h_{1c}}\,\zeta \,\left(\left(  - 2\,\alpha \,\zeta  \right)  + {\sqrt{{h_c}}}\,\alpha \,\zeta \,\Gamma
   \right)\nonumber\\
c_1&=&2\,\left({{a_{1}}}^2-1 \right) \,{h_c}\,\left(
{h_{1c}}\,{{x_c}}^2\,\left(2\,\alpha -1 \right)  +
\,\alpha \,\zeta\right)\\
c_2&=&8\,{{a_{1}}}^2\,b\,{{h_c}}^2\,{{x_c}}^2\,\alpha\nonumber\\
\end{eqnarray*}
where $h_{1c}\equiv dh/dx(x_c)$.
For a given position of the critical point $r_c$ equations ~({\ref{App_p1}) and ~({\ref{App_poly})
are used to calculate $p_c=(dv/dx)_c$ and $a_1=v_s/v_{c}$.
The mass-loss rate is straightforwardly calculated from
${\displaystyle {\dot M}=\frac{8\pi G M}{\sigma v_{th}}
\left(\frac{\Gamma k}{2\mu}\right)^{1/\alpha}}.
$

\begin{figure}
\centerline{\psfig{figure=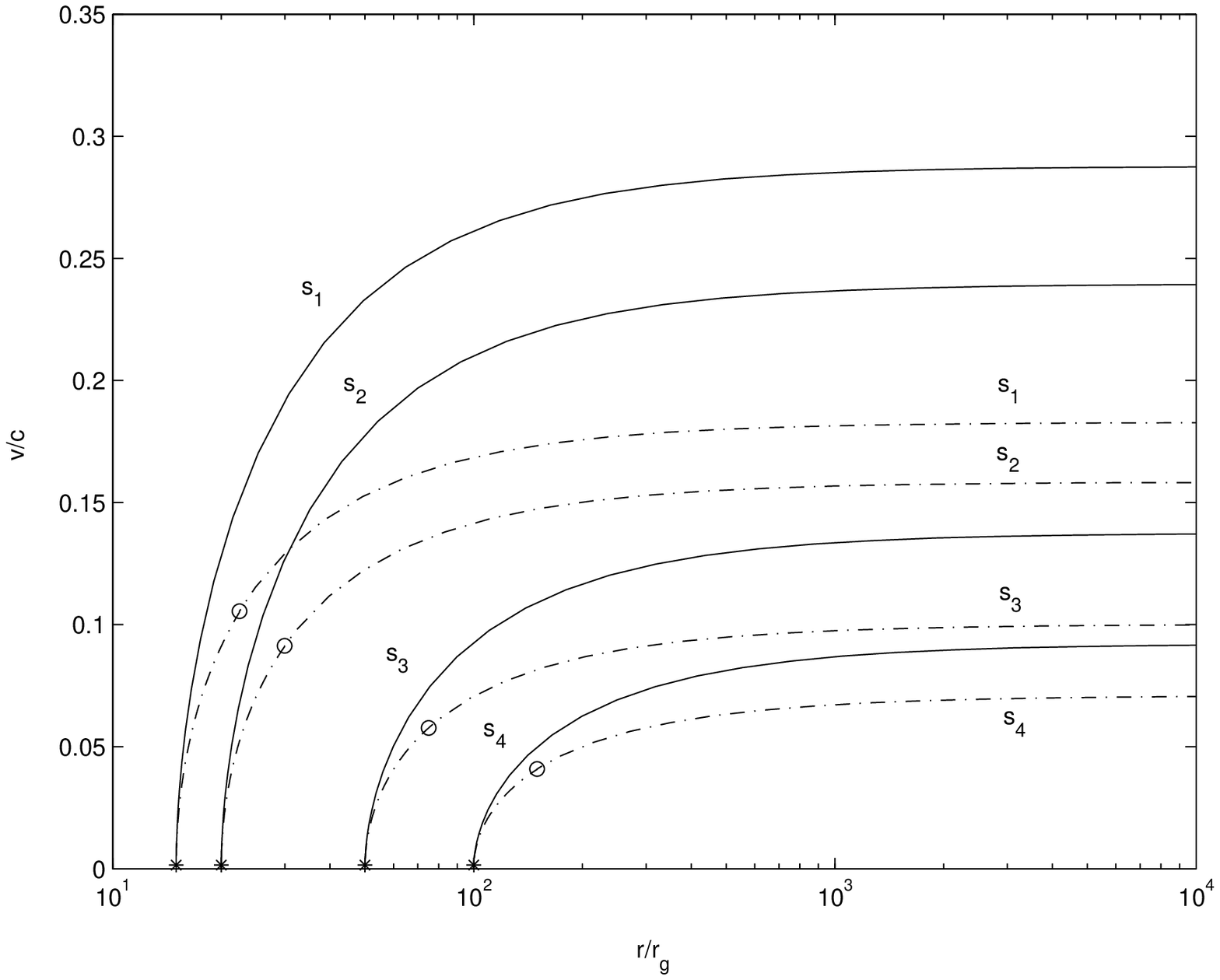,height=5in}} \caption{Solutions of the equation of
motion ~(\ref{Euler4}). Solid line - "Gravitationally Exposed Flow" (GEF) solution,
dashed line - standard line-driven wind (SLDW) regime. Stars indicate GEF critical
points, circles - SDLW critical points. cf. Figure 1. of Dorodnitsyn (2003) Labels
$s_1$ - $s_4$ mark solutions with different locations of the critical point.
The following set of parameters was adopted: $\Gamma=0.5$, $k=0.03$,
$M_{BH}=10^8 M_{\odot}$,
$T=4\cdot 10^4K$.}
\end{figure}

\end{document}